**T-matrix evaluation of three-dimensional acoustic radiation forces on nonspherical objects in Bessel beams with arbitrary order and location**

Zhixiong Gong[1,2,†], Philip L. Marston[2,‡], Wei Li[1,*]

[1]School of Naval Architecture and Ocean Engineering, Huazhong University of Science and Technology, Wuhan 430074, China

[2]Department of Physics and Astronomy, Washington State University, Pullman, Washington 99164-2814, United States

† Present address: Univ. Lille, CNRS, Centrale Lille, ISEN, Univ. Polytechniques Hauts-de-France, UMR 8520-IEMN, International laboratory LIA/LICS, F-59000 Lille, France

zhixiong.gong@iemn.fr

‡ marston@wsu.edu

* Corresponding author: hustliw@hust.edu.cn

Acoustic radiation forces (ARFs) induced by a single Bessel beam with arbitrary order and location on a nonspherical shape are studied using the T-matrix method (TMM) in three dimensions. Based on the radiation stress tensor approach and the multipole expansion method for the arbitrary Bessel beam, the ARF expressions are derived in terms of the incident and scattered beam shape coefficients independently with the corresponding homemade code packages. Several numerical experiments are conducted to verify the versatility of the TMM. The axial acoustic radiation forces (ARFs) of several typical shapes are considered in the analysis with the emphasis on the axial ARF reversal and the corresponding physical mechanism. This study may guide the experimental set-up to find negative axial ARFs quickly and effectively based on the predicted parameters with TMM. Relatively elongated shapes may be helpful for pulling forces in Bessel beams. Furthermore, the lateral ARFs for both convex and concave nonspherical shapes are also investigated with different topological charges, cone angles and offsets of the particle centroid to the beam axis in a broadband frequency regime. A brief theoretical derivation of the incident beam shape coefficients for the standing Bessel beams is also given. The present work could help to design the acoustic tweezers numerical toolbox which provides an acoustical alternative to the optical tweezers toolbox.

## 1 Introduction

Acoustic tweezers[1-4], an appropriate counterpart to optical tweezers[5], could be used for levitation[6,7], pulling forces[8-12], particle trapping[13,14], and even dynamic controls[15,16] in the fields of microfluidics and life sciences. Compared with optic tweezers, acoustic tweezers tend to exert a larger force over larger length scales with the same intensity since the radiation force is proportional to the ratio of the intensity to the velocity in the medium[2,17]. In general, there are two main schemes to design acoustic tweezers: the (quasi)standing wave scheme with dual beams[1,2,4] and the single beam structure[3]. Single-beam tweezers may be superior to general plane standing-wave tweezers in some respects, for instance, single-beam tweezers can continuously pull or push a particle over a large region because there are no multiple equilibrium positions[10,17]. Negative radiation force single-beam device could pull the target towards the source, which is of interest in both acoustical[8-12,17] and optical fields[18]. The physical mechanism is due to the asymmetric scattering of the incident fields on the target such that the scattering into the forward direction is relatively stronger than the scattering into the backward direction[8-10,18,19]. This is understood by the conservation of momentum and Newton's third law regarding reaction force between the acoustic field and the inside particle.[19,20]

Beams having the local properties of acoustic Bessel beams are candidates for single-beam tweezers which have been examined in theoretical[8-10,21] and experimental approaches[17,22]. The ordinary Bessel beam (OBB) possesses the axial maximum and azimuthal symmetry, while the helicoidal Bessel beams (HBBs) have an axial null and azimuthal phase gradient. Hefner and Marston conducted the experimental demonstration for the acoustical vortices by using simple four-panel piezoelectric transducers[23]. Recently, the transducer arrays[17,24], active spiral transducer[25,26], diffraction gratings[27] and metasurfaces[28] have been demonstrated to produce the local Bessel beams which coincide with the theoretical or simulation results. These fabrication technologies facilitate the experimental studies of Bessel beams and the possible applications in the fields of particle manipulations. In addition, the exact series solutions have been solved for the axial ARFs of *spherical* objects in an on-axis incident Bessel beam for both the ordinary[8,9] and helicoidal (vortex)[10,17] Bessel beams. The ARF produced by a Bessel vortex beam has also been studied via the optical theorem[29-33] which gives the relationship between the extinction and the scattering at the forward direction of the beam's plane wave components. However, it is still necessary to develop efficient and versatile numerical models to investigate the three-dimensional ARFs of particles with different shapes and other complicated conditions.

To this end, the T-matrix method (TMM) is introduced to the field of acoustic manipulations which has been demonstrated effective and very efficient for acoustic scattering[34] from spheroid[35-38], superspheroid[38,39] and finite cylinder with endcaps[40,41]. These nonspherical (convex or concave) shapes are very common to model the geometries of particles in biomedical engineering (e.g., cells and bacteria) and lab-chip technologies (e.g., drops with gravity and fibers) in the context of acoustophoresis, which could provide more accurate models and give a better prediction of the particle motions than the spherical shape. It is noteworthy that other numerical methods, such as the finite element method[42,43], boundary element method[44-46], finite volume method[47,48], lattice Boltzmann method[49,50] and finite-difference time-domain method[51], are also employed to successfully solve the acoustic radiation force of particles in fluid. Each numerical methods has its own metrics. However, for typical axisymmetric shapes, the TMM will be a superior approach since it will decrease the integral over the geometrical surface (three dimensions) into the line integral (two dimensions) and could save much computational cost, leading to high efficiency in the simulation experiments even at high frequencies[38,39]. Furthermore, based on the radiation stress tensor approach[52,29] and the independent derivations, the three-dimensional ARFs could be expressed and calculated in terms of the incident and scattered beam shape coefficients (see details below). Note that the transition matrix need only to calculate once for the scattered field and is independent of the incident beams, making the TMM even more efficient to calculate the forces on particles. It is convenient to obtain the incident beam shape coefficients for the ordinary and standing plane wave, however, may be challenging in the context of vortex beam since the particle may deviate the beam axis with an offset. For an ideal Bessel beam of arbitrary topological charge and location, general theoretical formulas of the incident beam shape coefficients are derived based on the multipole expansion method[35] and also given by Zhang[53].

It is important to investigate the particle dynamics in three dimensions with the on- or off-axis incidence since it could help the beam calibrate with the particle centroid in experimental setups with higher frequency regime. To some extent, the numerical method is an alternative to direct experimental approaches and more versatile than analytical investigations. In this paper, several numerical experiments are conducted based on the traditional T-matrix method with the emphasis on *nonspherical* objects which are typical in engineering practice and life sciences, such as the generalized superspheroid and finite cylinder with endcaps which may be used to model the biological cells or bacteria. This will extend the previous theoretical studies of the axial ARF[8-10] and numerical implementations of the off-axial ARF on a sphere[54,11] to cases of an object with complicate shapes placed in a Bessel beam with arbitrary location and order. Note that Few analytical solutions of the ARF on rigid spheroids have been derived in the long-wavelength limit[55,56], which has the potential for the Bessel beam illumination example. The physical mechanism of the axial ARF reversal for nonspherical shape is demonstrated by numerical experiments and the corresponding parameter conditions are discussed. In addition, the lateral and the axial ARFs for both convex and concave nonspherical shapes are discussed with emphasis on the dimensionless frequency, the cone angle of the Bessel beam and the offsets. The theoretical formulas of the Bessel standing wave with arbitrary orders and offsets are also briefly given with the numerical example.

## 2 Theoretical and numerical models

### 2.1 Radiation momentum stress tensor method

The ARFs comes from the transfers of linear momentum between the acoustic fields and particles which could be induced by the scattering or absorption. The radiation stress tensor approach[52,29] is widely employed to compute the static radiation force by integrating the time-averaged radiation stress tensor over a far-field spherical surface $S_0$. Consider a particle with arbitrary shape $S$ in an ideal fluid, as shown in Fig. 1, a Bessel beam is incident with arbitrary topological charge and location. Based on the momentum conservation, the radiation stress tensor $\mathbf{S}_T$ in the ideal fluid meets: $\nabla \cdot \langle \mathbf{S}_T \rangle = 0$. By integrating $\mathbf{S}_T$ over the particle surface $S$, the expression of the ARF could be written

as

$$\mathbf{F} = -\iint_S \langle \mathbf{S}_T \rangle \mathrm{d}\mathbf{S} = -\iint_S \langle L \rangle \mathrm{d}\mathbf{S} + \rho_0 \iint_S \mathrm{d}\mathbf{S} \cdot \langle \mathbf{uu} \rangle \quad (1)$$

where $\langle L \rangle$ is the time average of the Lagrangian density, $\rho_0$ is the density of the surrounding fluid and $\rho_0 \langle \mathbf{uu} \rangle$ described the average value of the flux of momentum density tensor. The integral in the ARF expression on the particle surface $S$ could be transferred to the far-field standard spherical surface $S_0$ according to the Gaussian theorem, having

$$\mathbf{F} = \iint_{S_0} \langle L \rangle \mathrm{d}\mathbf{S} - \rho_0 \iint_{S_0} \mathrm{d}\mathbf{S} \cdot \langle \mathbf{uu} \rangle \quad (2)$$

Note that Eq. (2) could apply for the arbitrary-shaped particle since the integral will, in fact, be conducted in a standard spherical surface, which is very important for both theoretical and numerical computations. For simplicity, the time average of the Lagrangian density is divided into three terms: the contribution of only the incident beam $\langle L_{ii} \rangle$, only the scattered field $\langle L_{ss} \rangle$, and the interaction of the incident and scattered fields $\langle L_{is} \rangle$, respectively,

$$\begin{aligned} \langle L_{ii} \rangle &= \frac{1}{2} \rho_0 \mathbf{u}_i \cdot \mathbf{u}_i - \frac{p_i^2}{2\rho_0 c_0^2} \\ \langle L_{ss} \rangle &= \frac{1}{2} \rho_0 \mathbf{u}_s \cdot \mathbf{u}_s - \frac{p_s^2}{2\rho_0 c_0^2} \\ \langle L_{is} \rangle &= \rho_0 \mathbf{u}_i \cdot \mathbf{u}_s - \frac{p_i p_s}{\rho_0 c_0^2} \end{aligned} \quad (3)$$

where $c_0$ is the velocity in the fluid, $\mathbf{u}_{i,s}$ are the incident and scattered velocity vectors with $\mathbf{u}_{i,s} = \mathbf{n} u_{i,s}$, $\mathbf{n}$ is the outward unit normal vector, $p_{i,s}$ are the first order pressure of the incident and scattered fields with the relationship $p_{i,s} = \rho_0 c_0 u_{i,s}$, leading to the vanish of $\langle L_{ss} \rangle$. In addition, there will be no transfer of linear momentum when the particle does not exist in the fluid, making the terms only related to the incident fields vanish

$$\iint_{S_0} \langle L_{ii} \rangle \mathrm{d}\mathbf{S} - \rho_0 \iint_{S_0} \mathrm{d}\mathbf{S} \cdot \langle \mathbf{u}_i \mathbf{u}_i \rangle = 0 \tag{4}$$

Inserting Eqs. (3) and (4) into (2), the ARF expression in terms of the velocity and pressure scalar quantities is

$$\mathbf{F} = -\iint_{S_0} \left\langle \frac{p_i p_s}{\rho_0 c_0^{\ 2}} \right\rangle \mathbf{n} \mathrm{d}S - \rho_0 \iint_{S_0} \langle u_s u_s \rangle \mathbf{n} \mathrm{d}S - \rho_0 \iint_{S_0} \langle u_s u_i \rangle \mathbf{n} \mathrm{d}S \tag{5}$$

By using the relations between the velocities (pressures) and complex velocity potentials for both the incident and far-field scattered fields, such as $\mathbf{u}_{i,s} = \nabla \Phi_{i,s}$, $p_{i,s} = i\omega\rho_0 \Phi_{i,s}$ and $u_{i,s} = \mathbf{n} \cdot \nabla \Phi_{i,s} = \partial \Phi_{i,s} / \partial r$ (far-field approximation), the expression of ARF in terms of velocity potentials could be

$$\mathbf{F} = \frac{1}{2} \rho_0 k^2 \iint_{S_0} \mathrm{Re} \left\{ \left( \frac{i}{k} \frac{\partial \Phi_i}{\partial r} - \Phi_i \right) \Phi_s^{\ *} - \Phi_s \Phi_s^{\ *} \right\} \mathbf{n} dS \tag{6}$$

where $k$ is the wave number, * denotes complex conjugation. $\Phi_i$, $\Phi_s$ denote the incident and scattered complex velocity potentials, Re means the real part of a complex number.

From the view of numerical computation, the TMM is an efficient tool to compute acoustic scattering on nonspherical objects. At present, this method will be further extended for ARFs which is closely related to the incident and scattered fields. In the TMM formulation, the velocity potentials of the incident and scattered fields could be

expanded as[35-41]

$$\Phi_i = \Phi_0 \sum_{nm} a_{nm} j_n\left(kr\right) Y_{nm}\left(\theta,\varphi\right) \quad (7)$$

$$\Phi_s = \Phi_0 \sum_{nm} s_{nm} h_n^{(1)}\left(kr\right) Y_{nm}\left(\theta,\varphi\right) \quad (8)$$

where $a_{nm}$ and $s_{nm}$ are the incident and scattered coefficients of expansion (others prefer to call the incident and scattered beam-shape coefficients), $\Phi_0$ is the beam amplitude, $j_n\left(kr\right)$ and $h_n^{(1)}\left(kr\right)$ are the spherical Bessel and Hankel functions of the first kind, respectively. In the far field ( $kr \to \infty$ ), the following asymptotic expressions of the spherical Bessel function and Hankel function of the first kind are used respectively, as $j_n\left(kr\right) \sim i^{-(n+1)} e^{ikr} \big/ 2kr + i^{n+1} e^{-ikr} \big/ 2kr$ and $h_n^{(1)}\left(kr\right) \sim i^{-(n+1)} e^{ikr} \big/ kr$ . $Y_{nm}\left(\theta,\varphi\right)$ denotes the normalized spherical harmonics. The transition relationship between $a_{nm}$ and $s_{nm}$ is given by $s_{nm} = T_{nm,n'm'} a_{n'm'}$ , where $T_{nm,n'm'}$ denotes the transition matrix which only depends on the properties of the object, including the geometrical shape, the material composition and the boundary conditions at the interface, and otherwise is independent of the sources. For the exact series solution, the transition matrix could be considered as $T = \left(s_n - 1\right) \big/ 2$ without dependence on the azimuthal index $m$ for spheres, which is in fact the partial-wave coefficients $a_n$ with the scattering coefficients $s_n$ known for a wide variety of spheres[57] and may be taken as a special case for the TMM. Varadan et al. also gave the explicit expressions of the transition matrixes for acoustic soft, hard and fluid sphere[58], which all coincide with those obtained with the exact series solutions. It is noteworthy that both the TMM[35-41] and the series solution for scattering by a sphere can be truncated at appropriate indices ( $N_{\max}$ ) in computations, which make the asymptotic expressions of scattered fields convergent in the far-field. After using the far-field asymptotic expressions for the scattered velocity potentials and implementing several algebraic manipulations (including the recursion relation of the spherical Bessel function and its derivative $j_n' = (n/kr) j_n - j_{n+1}$ with the variable $kr$), the ARF could be given briefly in terms of the incident and scattered beam shape coefficients, such that

$$\mathbf{F} = \frac{1}{2}\rho_0 k^2 \Phi_0^2 \iint_{S_0} \mathrm{Re}\left\{ -\sum_{nm}\sum_{n'm'} \frac{i^{n'-n}}{(kr)^2} (a_{nm} + s_{nm}) s_{n'm'}^{*} Y_{nm}(\theta,\varphi) Y_{n'm'}^{*}(\theta,\varphi) \right\} \mathbf{n} dS \tag{9}$$

which could be applied for radiation force with arbitrary orientation and agrees with Eqs. (7) and (9) in Silva's work[59]. The differential surface area is $\mathrm{d}S = r^2 \sin\theta d\theta d\varphi$. The dimensionless ARF $\mathbf{Y}$ is introduced to coincide with the exact solutions for spheres with the relationship as

$$\mathbf{F} = \pi r_0^2 I_0 c_0^{-1} \mathbf{Y} \tag{10}$$

where $I_0 = (\rho_0 c_0/2)(k\Phi_0)^2$ and $r_0$ is the characteristic dimension of the target. The outward unit normal vector is $\mathbf{n} = \sin\theta\cos\varphi \mathbf{e_x} + \sin\theta\sin\varphi \mathbf{e_y} + \cos\theta \mathbf{e_z}$ in Cartesian ordinates. Hence,

$$\begin{aligned}\mathbf{Y} = -\frac{1}{\pi (kr_0)^2} \iint_{S_0} \mathrm{Re}\left\{ \sum_{nm}\sum_{n'm'} \frac{i^{n'-n}}{(kr)^2} (a_{nm} + s_{nm}) s_{n'm'}^{*} Y_{nm}(\theta,\varphi) Y_{n'm'}^{*}(\theta,\varphi) \right\} \\ \times \left( \sin\theta\cos\varphi \mathbf{e_x} + \sin\theta\sin\varphi \mathbf{e_y} + \cos\theta \mathbf{e_z} \right) r^2 \sin\theta d\theta d\varphi \end{aligned} \tag{11}$$

The integration could be simplified easily by using Eqs. (15.150-152) in Ref. [60] for the integration involving the spherical harmonics and circular functions with the detailed derivations given in Appendix A. Finally, the axial and lateral components of the dimensionless ARFs could be derived independently as

$$Y_x = \frac{1}{2\pi (kr_0)^2} \mathrm{Im}\left\{ \sum_{nm} (a_{nm} + s_{nm}) \begin{pmatrix} -s_{n+1,m+1}{}^{*} b_{n+1,m} - s_{n-1,m+1}{}^{*} b_{n,-m-1} \\ +s_{n+1,m-1}{}^{*} b_{n+1,-m} + s_{n-1,m-1}{}^{*} b_{n,m-1} \end{pmatrix} \right\} \tag{12}$$

$$Y_y = \frac{1}{2\pi (kr_0)^2} \mathrm{Re}\left\{ \sum_{nm} (a_{nm} + s_{nm}) \begin{pmatrix} s_{n+1,m+1}{}^{*} b_{n+1,m} + s_{n-1,m+1}{}^{*} b_{n,-m-1} \\ +s_{n+1,m-1}{}^{*} b_{n+1,-m} + s_{n-1,m-1}{}^{*} b_{n,m-1} \end{pmatrix} \right\} \tag{13}$$

$$Y_z = \frac{1}{\pi (kr_0)^2} \mathrm{Im}\left\{ \sum_{nm} (a_{nm} + s_{nm}) \left( s_{n+1,m}{}^{*} c_{n+1,m} - s_{n-1,m}{}^{*} c_{nm} \right) \right\} \tag{14}$$

where the coefficients are $b_{n,m} = \left[ (n+m)(n+m+1)/(2n-1)(2n+1) \right]^{1/2}$ and $c_{n,m} = \left[ (n+m)(n-m)/(2n-1)(2n+1) \right]^{1/2}$. As observed from Eqs. (12-14), the three-dimensional ARFs could be obtained once the scattered beam shape coefficients are calculated from the incident coefficients through various methods, such as the partial wave series solution, the T-matrix method and other kinds of theoretical and numerical methods. It is noteworthy that the theoretical expressions of the axial ARF from spherical shapes in the zeroth- and first-order Bessel beams are derived in Refs. [8-10] and three-dimensional ARFs for arbitrarily located elastic sphere in Ref. [11].

**2.2 A brief review of incident beam shape coefficients of arbitrary Bessel beams**

Consider the Bessel beam is placed in an arbitrary location relative to the particle, as shown with the coordinate system in Fig. 2. The origin of the $Oxyz$ system $O$ coincides with the particle centroid, while the origin of the Bessel beam $O_B$ in the $O_B x' y' z'$ system coordinates is located at $(x_0, y_0, z_0)$ in $Oxyz$. The velocity potential of a Bessel beam with arbitrary topological charge $M$ and location could be expressed

as

$$\Phi_B = \Phi_0 i^M e^{ik_z(z-z_0)} J_M(k_r R') e^{iM\varphi'} \tag{15}$$

where $R' = \sqrt{(x-x_0)^2 + (y-y_0)^2}$ and $\varphi' = \tan^{-1}[(y-y_0)/(x-x_0)]$ in Fig. 2 describe the radius and azimuthal angle of the field point $(x, y, z)$ in the $O_B x' y' z'$ system. $k_z = k\cos\beta$ and $k_r = k\sin\beta$ are the axial and transverse component of the wave number $k = \omega/c_0$ with $\beta$ the cone angle of the Bessel beam and $\omega$ the angular frequency. By using the addition theorem for the Bessel functions and the exact solution to the integral on the hybrid product including the associated Legendre, Bessel, and exponential functions in spherical coordinates as

$$\int_{\theta=0}^{\pi} \mathrm{d}\theta \sin\theta e^{ikr\cos\beta\cos\theta} P_n^m(\cos\theta) J_m(kr\sin\beta\sin\theta) = 2i^{n-m} P_n^m(\cos\beta) j_n(kr) \tag{16}$$

The incident beam shape coefficients of the Bessel beam with arbitrary topological charge and location could be derived[35]

$$a_{nm} = 4\pi\xi_{nm} i^{n-m+M} P_n^m(\cos\beta) \times e^{-ik_z z_0} J_{m-M}(\sigma_0) e^{-i(m-M)\varphi_0} \tag{17}$$

with the normalized coefficients $\xi_{nm} = [(2n+1)(n-m)!]^{1/2} [4\pi(n+m)!]^{-1/2}$, $\sigma_0 = k_r R_0$, $R_0 = (x_0^2 + y_0^2)^{1/2}$ and $\varphi_0 = \tan^{-1}(y_0/x_0)$. When the offset is $(x_0, y_0, z_0) = (0,0,0)$, Eq. (17) will degenerate into the on-axis incidence situation of the Bessel beam.

The scattered beam shape coefficients $s_{nm}$ are still missing to calculate the three-dimensional ARFs based on Eqs. (12-14). In this work, a versatile TMM (which gives a linear relationship between the incident and scattered beam shape coefficients as $s_{nm} = T_{nm,n'm'} a_{n'm'}$) is introduced to the field of radiation forces. Note that the TMM in the acoustic field mainly considers the scattering field in underwater or elastic mediums instead of the further consideration on the acoustic radiation forces and torques. The present work is inspired to take advantage of this method (as discussed in the Introduction) which is very efficient for both spherical and aspherical shapes with rotational symmetry and only needs to be computed once for the transition (T) matrix as shown in the Appendix B. The TMM makes it possible to employ nonspherical shapes to model the real interesting particles in life science and engineering in an exacter manner.

## 3 Numerical results and discussion

### 3.1 Validation of T-matrix method for acoustic radiation force

To verify the correctness of the TMM, several examples are implemented for a rigid sphere in the ordinary (OBB, blue dashed line) and first-order helicoidal Bessel beams (FHBB, red solid line) as shown in Fig. 3(a). Both cases are under on-axis incidence. The axial ARFs $Y_z$ for the OBB case are extracted from Fig. 2 of Ref. [8] with the half-cone angle $\beta = 60^\circ$ [blue circles in Fig. 3(a)], while those for the HBB are extracted from Fig. 1 of Ref. [10] with $\beta = 66.42^\circ$ [red triangles in Fig.3(a)]. The reference results are calculated with the exact series solutions (partial-wave series method). As shown in Fig. 3(a), all the TMM results agree well with the series solutions. In addition, the axial ARFs $Y_z$ for a rigid sphere located off the OBB axis have been calculated by TMM and partial-wave series method based on the multipole expansion method, with the references given by the boundary element method (Fig. 11 in Ref. [46]). The cone angle is $\beta$=30$^\circ$, dimensionless frequency $ka = 1$, and $x_0$ describes the offset with the length unit in meters. As shown in Fig. 3(b), the results from the TMM and partial-wave series coincide with each other and agree well with those from the Boundary element method. Moreover, the TMM has been demonstrated for the scattering from spheroid[36-39], finite cylinder[40,41] in plane wave and Bessel beams, and hence it could be applied for the radiation forces for these shapes convincingly provided that the incident and scattered coefficients of expansion are computed correctly [see Eqs. (12-14)]. It is noteworthy that the incident and scattered coefficients (occurring as column vectors in our numerical computations with the TMM) will be assembled in the same way for the radiation force. This will further verify the effectiveness of the present derivation of the ARF using the TMM and the corresponding homemade codes.

Furthermore, the convergence curves of the TMM are discussed in terms of the axial ARF versus different truncation number $N_{\max}$ for a biconcave shape [see Fig. 3(c) for the 2D schematic, taking the "peanut-shaped" generalized superspheroid as an example]. The definitions of $a$ and $b$ for the generalized superspheroid in Fig. 3(c) are analogous with those for a spheroid[36]. For the rotational symmetry of the generalized superspheroid, the distance of the surface field to the origin (center of the object) could be expressed as $r_S(\theta) = \left(a^2\cos^2\theta + b^2\sin^2\theta\right)^{1/2}$, where $\theta$ is the polar angle of the surface field point, which are used by implementing an integral over the object surface to obtain the transition matrix $T_{nm,n'm'}$ relating the incident coefficients to the scattered coefficients of expansion. Note that due to the rotational symmetry, the integral involving the term $r_S(\theta)$ over the surface is only dependent on the polar angle, otherwise independent of the azimuthal angle. Fig. 3(d) depicts the axial ARFs of a rigid "peanut-shaped" generalized superspheroid with the aspect ratio $a/b = 4$. The incident wave is a first-order Bessel beam with an arbitrary cone angle (here we choose $\beta = 66.42^\circ$) and the dimensionless frequency $kr_0$=8 where $r_0$ is the characteristic length of the nonspherical object. Under this circumstance, $r_0$ is the larger value between $a$ and $b$. The on-axis incidence is described by the blue solid line with triangles, while the off-axis case with the offset $(x_0, y_0) = (0.1\pi/kr_0, 0.1\pi/kr_0)$ is described by the red solid line with circles. $(x_0, y_0)$ are the translational coordinates of the beam center with respect to that of the object (i.e. the origin of the considered coordinates). A convergence test for a rigid finite cylinder with spherical endcaps has been conducted in Ref. [40] and hence omitted for brevity. All these curves converge very fast versus $N_{\max}$, which further demonstrates the efficiency of the present TMM for ARF. In the following computations, the truncation number is set as $N_{\max} = 2 + \mathrm{Int}\left(8 + kr_0 + 4.05\sqrt[3]{kr_0}\right)$, which could ensure the accuracy and convergence of the present computations according to our tests. The symbol Int means to round the following number towards the positive infinity. Both the accuracy and convergence tests and related discussions provide enough validation of the T-matrix method for acoustic radiation force.

### 3.2 Axial ARF reversal and physical mechanism

Two numerical experiments were conducted with the emphasis on the negative ARFs exerted on the rigid oblate and prolate spheroids in Bessel beams and the related physical mechanisms. The Neumann boundary condition was applied throughout for objects including spheroids, generalized superspheroids and finite cylinders in the following. In this Section, the rotational axis of the rigid spheroid coincides with the incident beam axis. Note that the relative orientation of the spheroid to the beam axis changes the scattering field from the particle, leading to the alteration of the transfer of linear momentum from the incident acoustic beam to the particle. The $Y_z$ of the oblate and prolate spheroids versus the dimensionless frequency $kr_0$ in the FHBB are depicted in Fig. 4(a) for $a/b=1/2$ and Fig. 4(b) for $a/b=2$, with $\beta=30^\circ$, $66.42^\circ$, and $80^\circ$. $a$ is the polar radius and $b$ is the equatorial radius.[29] The ranges including the negative ARFs in panels (a) and (b) are zoomed in and presented in panels (c) and (d) of Fig. 4, respectively. It implies that a large $\beta$ (sufficiently nonparaxial) may facilitate the pulling force since the negative ARFs appear for both cases with $\beta=80^\circ$ in the considered region, while it fails for $\beta=30^\circ$. Especially, negative ARF is impossible for plane waves ($\beta=0^\circ$) with passive spheres[19,61]. The term in Eq. (21) of Ref. [19] ($F_z=P_{sca}c^{-1}\left(\cos\beta-\langle\cos\theta_s\rangle\right)$ without absorption) including $\cos\beta$ represents the momentum removed from the incident Bessel beam (which induces positive ARF) and the term including $-\langle\cos\theta_s\rangle$ gives the axial projection of the momentum transport associated with the scattered field (which may induce positive or negative ARF), where $\theta_s$ is the polar angle of the field point with respect to the positive *z*-direction [see Fig. 3(c)]. The schematic of Fig. 1 qualitatively describes how to produce a negative ARF on an arbitrary object. The red solid arrows in the forward hemisphere denote the total scattered fields with the forward axial components, while the blue dashed arrows in the backward hemisphere represent the total scattered field with the backward axial components. The resultant forward component of all the scattering denoted by the red solid arrows are relatively larger than the resultant backward component of all scattering denoted by the blue dashed arrows. To further reveal the physical mechanism quantitatively, the angular dependences of the scattered form functions versus the scattered polar angle $\theta_s$ for the oblate spheroid in the FHBB with $\beta$=80° are plotted in Fig. 4(e) with $kr_0 = 1.8$ [marked as the red pentagram in Fig. 4(c)] and $kr_0 = 2.1$ [marked as the blue pentagram in Fig. 4(c)]. The black dotted line denotes the direction of the incident wave vector (i.e. $\theta_s = \beta$). As shown in the enlarged view in Fig. 4(c), the ARF is negative at $kr_0 = 1.8$, and otherwise positive at $kr_0 = 2.1$. It can be observed in Fig. 4(e) that for $kr_0 = 1.8$, the scattering dominates in the forward directions with $\theta_s < \beta$, resulting in the negative ARF; for $kr_0 = 2.1$, the scattering in the backward is relatively stronger than that in the forward, leading to the positive ARF. To better understand the relationship between the axial ARFs and two-dimensional scattering patterns, the scattering patterns of form functions for both the oblate and prolate spheroids in the first-order HBB (*M*=1) with cone angle $\beta$=80° are given versus different dimensionless frequencies ranging from $kr_0$ =10 to $kr_0$ =0.2 [See Mov1 in Supplemental Material at URL]. Note that at high frequencies, the forward scattering is comparable with the scattering in the backward hemisphere, however, the positive axial force induced by the incident wave ( $P_{sca}c^{-1}\cos\beta$ ) is important and lead to the resultant axial force positive.

### 3.3 Pulling forces on typical nonspherical objects

After giving an explicit explanation of the physical mechanism for the negative ARF, the emphasis will be put on the parameter conditions for exerting the pulling force on several typical objects, which may have potential applications in acoustophoresis, surface chemistry, atomic physics, ultrasonic medicine, reduced gravity environment, and so on. Panels (a-d) of Fig.5 study the influence of the topological charges (orders) of the Bessel beams for a "peanut-shaped" generalized superspheroid with $a/b=2$ for an on-axis incidence. The 2D plots depict the negative ARF “islands” in the $\left(kr_0,\beta\right)$ domain and the white domains stand for the positive ARFs (not shown numerically). The islands of the negative ARF are different between the OBB and HBBs since panel (a) has two subregions, while panels (b-d) have one subregion under consideration. For the HBBs, the frequencies of the negative ARF seem to increase with the increase of the beam order. To discuss the parameter of the aspect ratio, the 2D plots of a generalized superspheroid with $a/b=3$ and $a/b=4$ in the FHBB are given in panel (e) and (f) [compared with panel (b)], respectively. These results imply that the distributions of the negative ARF depend on both the beams and objects. However, the central frequencies do not change greatly with the aspect ratios. The oblate case for the generalized superspheroid is also described in panel (g). This shape is like some biological cells, the red blood cell with a dip in the center for example. In biomedicine or reduced gravity environment, the finite cylinder shapes with endcaps are helpful to model several kinds of bacteria or space shuttle, which will be discussed as follows. Capsule-shaped (cylinder with spherical endcaps[40,41]) objects are investigated for both the on-axis [panel (h)] and off-axis incidences [panels (i-l)]. The aspect ratio is $l/b = 2$ for all the cases, where $l$ is the half length of the total cylinder and $b$ is the radius of the cylindrical portion.[40] Panels (h-j) are for FHBB while Panels (k,l) are for SHBB. The beam axis is shifted off the axis of the object in the transverse plane as $\left(0.1\pi/kr_0, 0.1\pi/kr_0\right)$ for Fig. 5(i,k) and $\left(0.5\pi/kr_0, 0.5\pi/kr_0\right)$ for Fig. 5(j, l). Note that there is no need for the extra computational cost for the off-axis incidence compared with the on-axis case[35]. By comparison, the area of the negative force island decreases with a larger offset with respect to the object's center for both the FHBB and SHBB. It also implies that the negative ARFs occur at higher frequencies with a larger offset by comparing Fig.5(i) with Fig.5(j) [or Fig.5(k) with Fig.5(l)]. Unfortunately, quantitative results for the orientation dependence of the negative force are not available although it is known that to induce a negative axial ARF the scattering in the forward hemisphere needs to be stronger than scattering in the backward hemisphere[29].

One of the most important results concerns the extent of the first-order HBB negative ARF regions in evident in Fig. 5 (b) and (h) where the cone angle $\beta$ can be as small as 54 degrees. For fixed rigid spheres on the axis of a first-order HBB it has long been known that conditions can be found giving negative axial ARF[8,10]. The smallest value of $\beta$ for a first-order HBB to produce negative ARF on a rigid sphere is known to be $\beta$ of approximately 63 degrees. Inspection of Fig. 5 (b) and (h) shows that for an appropriately elongated generalized superspheroid [Fig. 5(b)] and capsule [Fig. 5(h)] the associated values can be as small as 54 and 55 degrees, respectively. This suggests those shapes of elongated objects can be especially favorable for producing negative axial ARF in first-order HBB.

### 3.4 Three dimensional ARFs for typical nonspherical objects

The three-dimensional ARFs (from the first to third column) of typical concave and convex shapes are investigated versus the dimensionless frequencies $kr_0$, the cone angle of the Bessel beam $\beta$ and the transverse offset $(x_0, y_0)$, as shown in Fig. 6. The first-order HBB is considered with both on- (first row) and off-axis incidences (second and third rows). The offset is set as $(x_0, \mathrm{y}_0) = (0.5\pi/kr_0, 0.5\pi/kr_0)$. The generalized superspheroid with $a/b = 2$ (first and second rows) is discussed at first and it could be found that the transverse ARFs vanish for the on-axis incidence because of the rotational symmetry of both the incident Bessel beam and geometric shape. However, for the off-axis incidence, the transverse ARFs will exist, see Fig. 6(d) and 6(e). To investigate the effect of the geometric shapes on the three-dimensional ARFs, the ARFs of a smoothed spheroid with the same aspect ratio $a/b = 2$ and offset are given in the third row of Fig. 6. By comparison of the ARFs in the second and third rows, the main profiles of ARFs in the two-dimensional $(kr_0, \beta)$ regions are similar for the same aspect ratio and off-set. However, there are some “jumps” in the ARFs patterns versus $(kr_0, \beta)$ at relatively high frequencies (e.g., $kr_0 \geq 4.75$). This is due to the fact that the scattering patterns will be more easily influenced when the aspect of the geometric shape comparable with the wavelength (i.e., relatively high frequency), which is further demonstrated by the similar ARFs patterns at low frequencies, see the second and third rows.

In addition, the three-dimensional ARFs of the generalized superspheroid with $a/b = 2$ are studied versus the offset $x_0$ and $y_0$ in ordinary (order=0), first-order (order=1) and second- order (order=2) Bessel beams, as depicted in Fig. 7. The range of the offset is $-1 \leq x_0 \leq 1$ and $-1 \leq y_0 \leq 1$, with the increase are $\Delta x_0 (\Delta y_0) = 0.02$. The incident dimensionless frequency is $kr_0 = 10$ and cone angle $\beta = 30^\circ$. As observed in the first ($Y_x$) and second ($Y_y$) columns, the transverse ARFs versus the offsets $(x_0, y_0)$ have the rotational symmetry with the angle of $\pi/2$. This could be easily understood by the reciprocity of transverse ARFs in $x$ and $y$ directions. Furthermore, the axial ARFs of OBB and HBBs are different such that the maximum value occurs at the axis for the OBB while at the concentric ring for the HBBs, which depends on the structure profiles of the Bessel beams. The three-dimensional ARFs could be used to discuss the trapping stability and the dynamic motions (axial translocation and orbital rotation around the beam axis) of particles in the Bessel beams. To further understand the three-dimensional ARFs of the generalized superspheroid versus different dimensional frequencies and transverse offsets, the ARFs are given when the superspheroid is placed in the ordinary ($M$=0), first-order ($M$=1), and second-order ($M$=2) Bessel beams, respectively, with a fixed cone angle $\beta = 30^\circ$ and the frequencies ranging from $kr_0$=0.2 to $kr_0$=10 [see Mov2 in Supplemental Material at URL]. The transverse forces are given in the form of arrow patterns, while the axial force is placed in the background with the colormaps.

### 3.5 Axial ARFs of rigid spheres in Standing Bessel Waves

The incident beam shape coefficients of a traveling Bessel beam are given theoretically in Eq. (17)[35], which could be easily extended for the standing Bessel waves. The velocity potential $\Phi_{SB}$ of a standing Bessel beams could be written as

$$\Phi_{SB} = \Phi_0 i^M \left[ Ae^{ik_z(z-z_0+h)} + Be^{-ik_z(z-z_0+h)} \right] \times J_M\left(k_r R'\right) e^{iM\varphi'} \tag{18}$$

where $h$ is the axial distance between the particle centroid and nearest pressure antinode, $A$ and $B$ are the amplitudes of the two beams with opposite propagation. To keep the energy of the standing fields the same as the traveling Bessel beam, one has $A^2 + B^2 = 1$, with $B = \alpha A$ and $A = 1\big/\sqrt{1+\alpha^2}$. Conducting the similar derivation of the beam shape coefficients of a single Bessel beam, the theoretical derivations for the standing Bessel beams could be de obtained as

$$a_{nm} = 4\pi\xi_{nm}\left[Ae^{ik_z(-z_0+h)} + Be^{-ik_z(-z_0+h)}(-1)^{n-m}\right]i^{n-m+M}P_n^m(\cos\beta)J_{m-M}(\sigma_0)e^{-i(m-M)\varphi_0} \quad (19)$$

Both the axial ARFs of a rigid sphere in a standing ($\alpha = 1$) and traveling ($\alpha = 0$) Bessel beams with different orders (order=0, 1, 2, and 3) are calculated versus the dimensional frequency $ka$ with a fixed cone angle $\beta = 60^\circ$. As observed in Fig. 8, axial force curves in standing Bessel waves show intuitive oscillation characteristics versus the dimensionless frequency as similar to the plane standing waves, which are different from those in traveling Bessel beams. However, the fabrication set-ups for Bessel beams and calibration of two counter-propagating Bessel beams will be challenging in experimental and applied investigations.

## 4 Conclusions

Computation of three-dimensional acoustic radiation forces on objects with complex geometrical shapes and boundary conditions is a challenging topic in engineering applications. Previous derivations of the acoustic radiation pressure are based on the long wavelength approximation[62-64] which has the limitation that the particle size is much smaller than the acoustic wavelength. Recently, the partial wave series solution has been introduced to study the ARF in the context of a Bessel beam without the limitation of computational frequencies. However, this exact solution may be restricted to certain shapes[8-10]. The T-matrix method is quite helpful for typical objects in engineering and especially efficient for shapes with rotational symmetry. In addition, the TMM could be employed for scattering problems involved in a waveguide[65] or multiple scattering[66] (including objects of multilayers and/or arbitrary numbers), which could be further extended for the ARFs based on the present work. The present numerical experiments demonstrate the effectiveness of the TMM to calculate the ARFs for several typical shapes, and the negative axial ARFs are obtained under certain conditions with the corresponding physical mechanisms. The TMM is very versatile for both spherical and nonspherical shapes with different material composition[34-41,58] once the geometrical shape functions could be given explicitly, providing an alternative to theoretical and experimental approaches. Other numerical methods, such as the finite volume method (FVM)[47,48], the (modified) finite element method (FEM)[68,69], the boundary element method (BEM)[44-46], the finite-difference time-domain method (FDTD)[51], and methods based on the ray acoustics approach[69] and the perturbation theory[70], may combine with the present derivation to provide more choices for the computations of ARF in Bessel beams. The TMM can be also used to calculate the acoustic radiation torques[71], which has been implemented in optics with the TMM[72] for a Gaussian beam incidence[73] by using the sums of products of the expansion coefficients for the integrals of the angular momentum fluxes[74]. The design of the acoustic tweezers numerical toolbox will benefit from the present work as similar to that in Optics[73]. It is anticipated that the three-dimensional ARFs could be obtained immediately once the scattered coefficients could be calculated according to a certain incident wave. The dynamic motions could be obtained for the axial translocation and orbital rotation around the beam axis. The long nonspherical shape may be especially favorable for producing negative axial ARF in Bessel beams, which is potential in the fields of microfluidics and life sciences. A brief theoretical derivation of the beam shape coefficients for the standing Bessel beams is given with the comparisons of axial ARFs in both standing and traveling beams, which may provide more possibility for the particle manipulations with vortex beams.

## Appendix:

### A. Detailed derivations of Three-dimensional ARFs

To conduct the integrals of products including spherical harmonics and trigonometric functions over the solid angle in Eq. (11), the following formulas should be introduced based on the Eqs. (15.150-152) in Ref. [60] for the three-dimensional dimensionless ARFs $Y_x$, $Y_y$ and $Y_z$, respectively

$$\begin{aligned}&\int_0^{2\pi}\int_0^{\pi} Y_{nm}(\theta,\varphi) Y_{n'm'}^{*}(\theta,\varphi)\sin^2\theta\cos\varphi\,\mathrm{d}\theta\,\mathrm{d}\varphi \\ &= -\frac{1}{2}\left(b_{n+1,m}\delta_{m',m+1}\delta_{n',n+1} - b_{n,-m-1}\delta_{m',m+1}\delta_{n',n-1} - b_{n+1,-m}\delta_{m',m-1}\delta_{n',n+1} + b_{n,m-1}\delta_{m',m-1}\delta_{n',n-1}\right);\end{aligned} \tag{A1}$$

$$\begin{aligned}&\int_0^{2\pi}\int_0^{\pi} Y_{nm}(\theta,\varphi) Y_{n'm'}^{*}(\theta,\varphi)\sin^2\theta\sin\varphi\,\mathrm{d}\theta\,\mathrm{d}\varphi \\ &= \frac{i}{2}\left(b_{n+1,m}\delta_{m',m+1}\delta_{n',n+1} - b_{n,-m-1}\delta_{m',m+1}\delta_{n',n-1} + b_{n+1,-m}\delta_{m',m-1}\delta_{n',n+1} - b_{n,m-1}\delta_{m',m-1}\delta_{n',n-1}\right);\end{aligned} \tag{A2}$$

$$\int_0^{2\pi}\int_0^{\pi} Y_{nm}(\theta,\varphi) Y_{n'm'}^{*}(\theta,\varphi)\sin\theta\cos\theta\,\mathrm{d}\theta\,\mathrm{d}\varphi = \left(c_{n+1,m}\delta_{m',m}\delta_{n',n+1} + c_{n,m}\delta_{m',m}\delta_{n',n-1}\right); \tag{A3}$$

where $\delta$ is the Kronecker delta function. Substituting Eqs. (A1)-(A3) into the three components into Eq. (11), the explicit expressions are obtained as Eqs. (12-14) with the corresponding coefficients therein.

### B. Explicit expression of transition (T) matrix

The incident and scattered beam shape coefficients are related by the transition matrix, depending on the geometric shape, material composition, and boundary conditions at the interface of the particle. For a rigid particle with rotational symmetry, the T matrix could be calculated as $\mathbf{T} = -\operatorname{Re}\mathbf{Q}\mathbf{Q}^{-1}$ with the element of the $\mathbf{Q}$ matrix derived as[36]

$$Q_{nm,n'm'}^{\sigma\sigma'} = \int_0^{\pi} \xi_{n'm'} j_{n'}\left(kr\right) P_{n'}^{m'}\left(\cos\theta\right) \xi_{nm} \left[ \frac{\partial h_n^{(1)}\left(kr\right)}{\partial r} P_n^m\left(\cos\theta\right) - \frac{r_\theta}{r^2} h_n^{(1)}\left(kr\right) \frac{\partial P_n^m\left(\cos\theta\right)}{\partial\theta} \right] \times r^2 \sin\theta \mathrm{d}\theta \int_0^{2\pi} \begin{pmatrix} \cos m'\varphi \\ \sin m'\varphi \end{pmatrix} \begin{pmatrix} \cos m\varphi \\ \sin m\varphi \end{pmatrix} \mathrm{d}\varphi \quad \text{(A4)}$$

where $r\left(\theta\right)$ is the geometric shape function and $r_\theta = \mathrm{d}r/\mathrm{d}\theta$ is the derivate of $r\left(\theta\right)$ with respect to the polar angle $\theta$ on the particle surface. Further details and simplified methods could be found in Refs. [36,40]. In fact, the $\mathbf{Q}$ matrix could be calculated for an arbitrary shape from the point of view of theory, however, there can be severe numerical difficulties in the general situation. The TMM is quite efficient for rotational shapes as demonstrated in the literature over the past decades.

## Acknowledgments

Work supported by HUST Postgraduate Overseas Short-term Study Program Scholarship and China Scholarship Council (Z.G.), NSFC (W.L. and Z.G.) and ONR (P.L.M).

## Figure captions

**Fig. 1** (Color online) Schematic of an arbitrary 3D object in an ideal fluid illuminated by a helicoidal Bessel beam (HBB) with arbitrary order and location. The acoustic scattering in the forward half-space (red solid arrows) is relatively stronger than the scattering in the backward half-space (blue dashed arrows), leading to a negative ARF.

**Fig. 2** (Color online) The coordinates relationship of the particle centroid [ $O(0,0,0)$ ] and beam origin [ $O_B(x_0, y_0, z_0)$ ]. $(x, y, z)$ is an arbitrary field point.

**Fig. 3** (Color online) Validations of the Axial ARFs calculated using the TMM compared with those from the exact solutions for a rigid sphere (*a*=*b*) in a Bessel beam based on the multipole expansion method. (a) On-axis incidence for with the orders *M*=0 (blue dashed line) and *M*=1 (red solid line). (b) Off-axis incidences with the order *M*=0 and cone angle $\beta=30^\circ$. (c) 2D schematic of a "peanut-shaped" generalized superspheroid placed on the axis of a Bessel beam. (d) Convergence tests for a rigid generalized superspheroid with aspect ratio $a/b=4$ for on-axis and off-axis incidences with the order of Bessel beam *M*=1.

**Fig. 4** (Color online) (a) The axial ARF of the rigid oblate spheroid with aspect ratio $a/b=1/2$ with $r_0=b$ ( $a<b$ ). The order of the Bessel beam is *M*=1 with cone angles

$\beta = 30^\circ$, $66.42^\circ$, and $80^\circ$. (b) Like panel (a) except that the particle is the prolate spheroid with $a/b = 2$ and $r_0 = a$ ($a > b$). Panels (c) and (d) depict the enlarged view of the negative ARF region for the rigid oblate and prolate spheroids, respectively. (e) Angular dependence of the scattered form functions versus the scattered polar angle $\theta_s$ for the oblate spheroid in the first-order HBB with $\beta$=80$^\circ$ for $kr_0 = 1.8$ [red solid line, corresponding to the red solid pentagram in (c)] and $kr_0 = 2.1$ [blue dash line, corresponding to the blue solid pentagram in (c)]. The black dotted line denotes the direction of the incident wave vector with $\theta_s = \beta$. Negative axial ARFs exist when the scattering in the forward hemisphere is relatively stronger than that in the backward hemisphere.

**Fig. 5** (Color online) Negative axial ARF "islands". The 2D plots depict only the negative ARFs in the $(kr_0, \beta)$ domain with colors, while the white domain stands for the non-negative ARFs. (a) the rigid generalized superspheroid with $a/b = 2$ under the on-axis incidence of the OBB (*M*=1). (b) Like panel (a) except that order=1. (c) Like panel (a) except that order=2. (d) Like panel (a) except that order=3. (e) Like panel (b) except that $a/b = 3$. (f) Like panel (b) except that $a/b = 4$. (g) Like panel (c) except that $a/b = 1/2$. $r_0 = b$ since $a < b$ in this case. This shape may model a red blood cell shape with a dip in the center. Panels (h-l): The 2D negative ARF islands for a capsule shape ($l$=2$b$) with $r_0 = l$. (h-j) are for the first-order (*M*=1) HBB with (h) on-axis and (i,j)

off-axis incidence. The offsets $\left(x_0, y_0\right)$ in the unit of meters are $\left(0,0\right)$, $\left(0.1\pi/kr_0, 0.1\pi/kr_0\right)$ and $\left(0.5\pi/kr_0, 0.5\pi/kr_0\right)$, respectively for (h-j). (k) Like panel (i) except that order=2. (l) Like panel (j) except that order=2.

**Fig. 6** (Color online) Three-dimensional ARFs versus dimensional frequency $kr_0$ and cone angle $\beta$. The first two columns describe the transverse ARFs while the third is the axial ARFs. A rigid generalized superspheroid with aspect ratio $a/b = 2$ under the (a-c) on-axis and (d-f) off-axis incidences of the first-order (*M*=1) HBB. The offset is set as $\left(x_0, \mathrm{y}_0\right) = \left(0.5\pi/kr_0, 0.5\pi/kr_0\right)$. The third row is as same as the second except that the particle is a smoothed rigid spheroid with $a/b = 2$.

**Fig. 7** (Color online) Three-dimensional ARFs (first to third columns: $Y_x$, $Y_y$, and $Y_z$) versus transverse offset $x_0$ and $y_0$ for Bessel beam with different orders: (a-c) OBB (*M*=0); (d-f) first-order (*M*=1) HBB; (g-i) second-order (*M*=2) HBB.

**Fig. 8** (Color online) Axial ARFs of a rigid sphere in a standing or traveling Bessel beam with a fixed cone angle $\beta = 60^\circ$.

**Fig. 1**

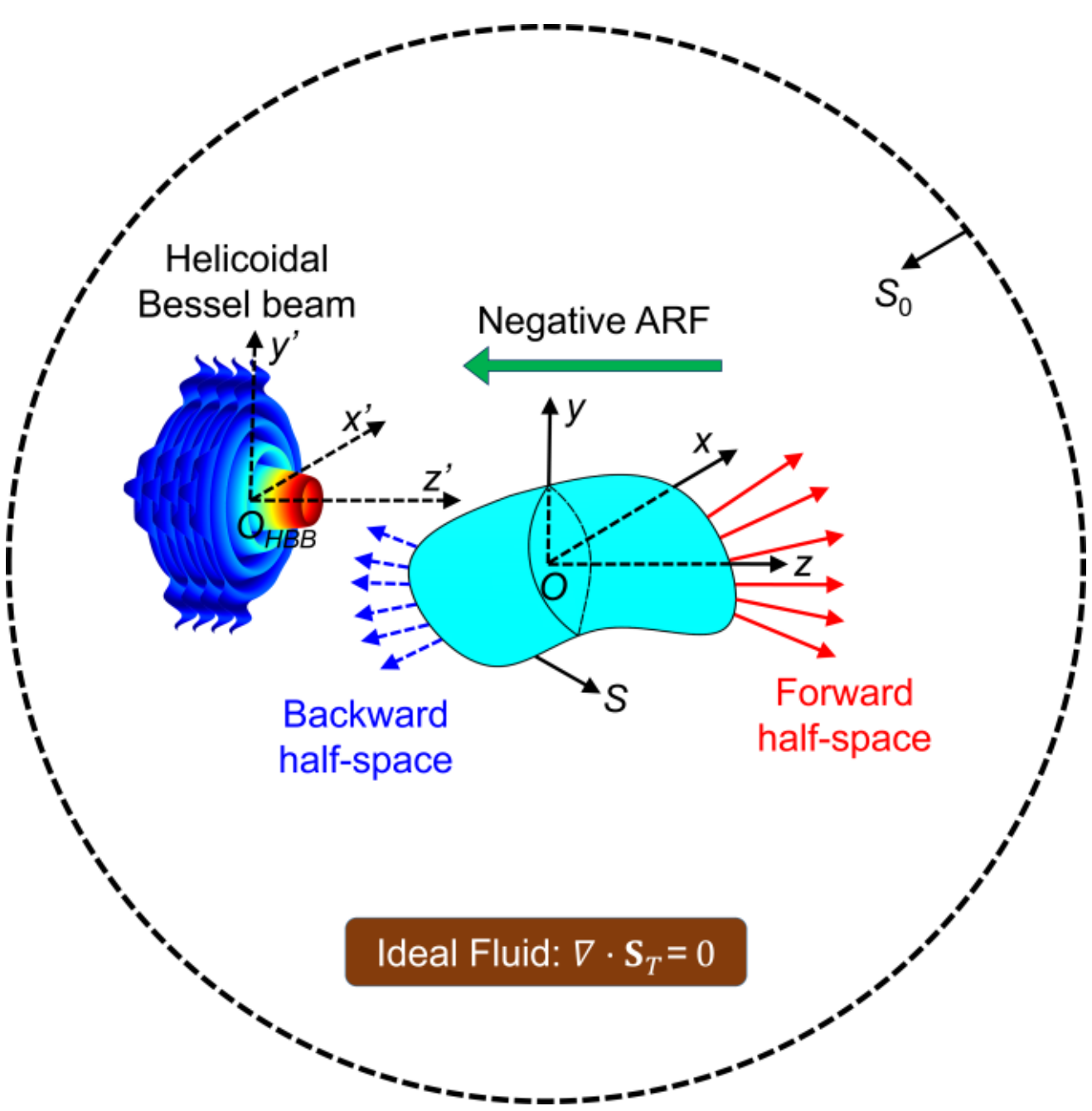

**Fig. 2**

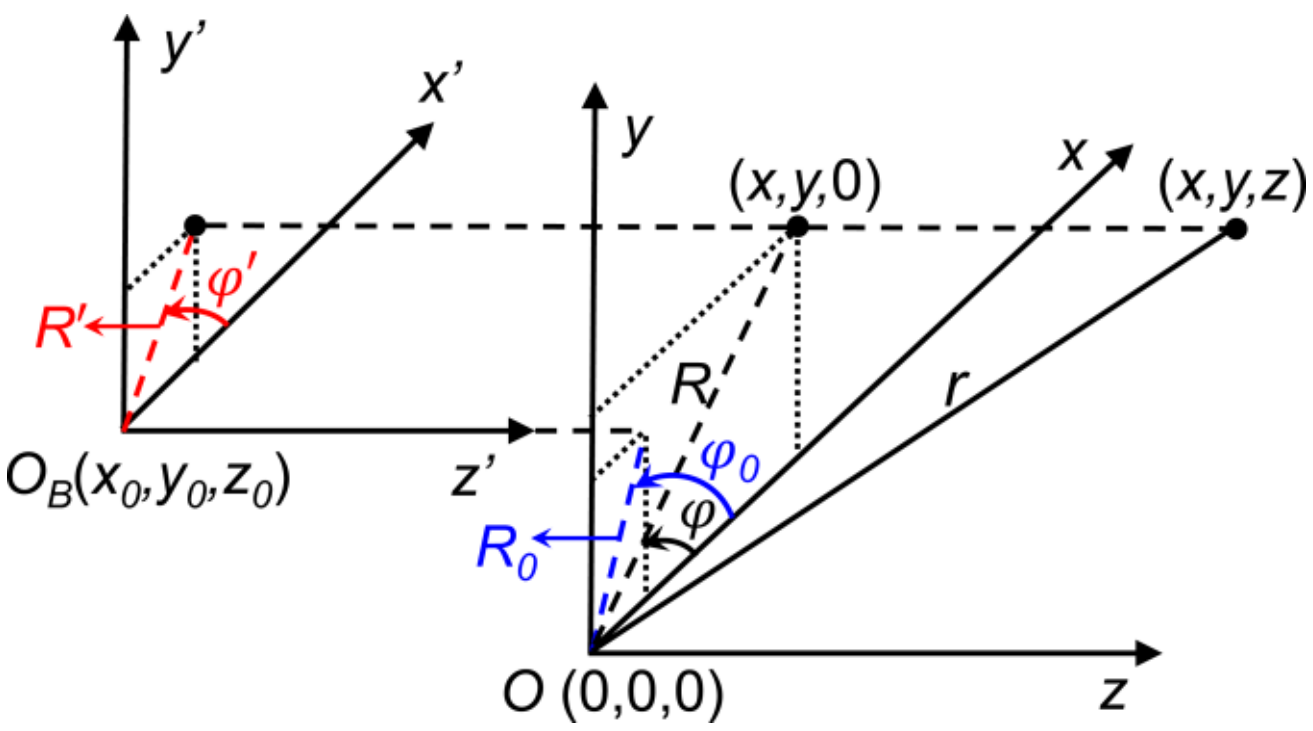

**Fig. 3**

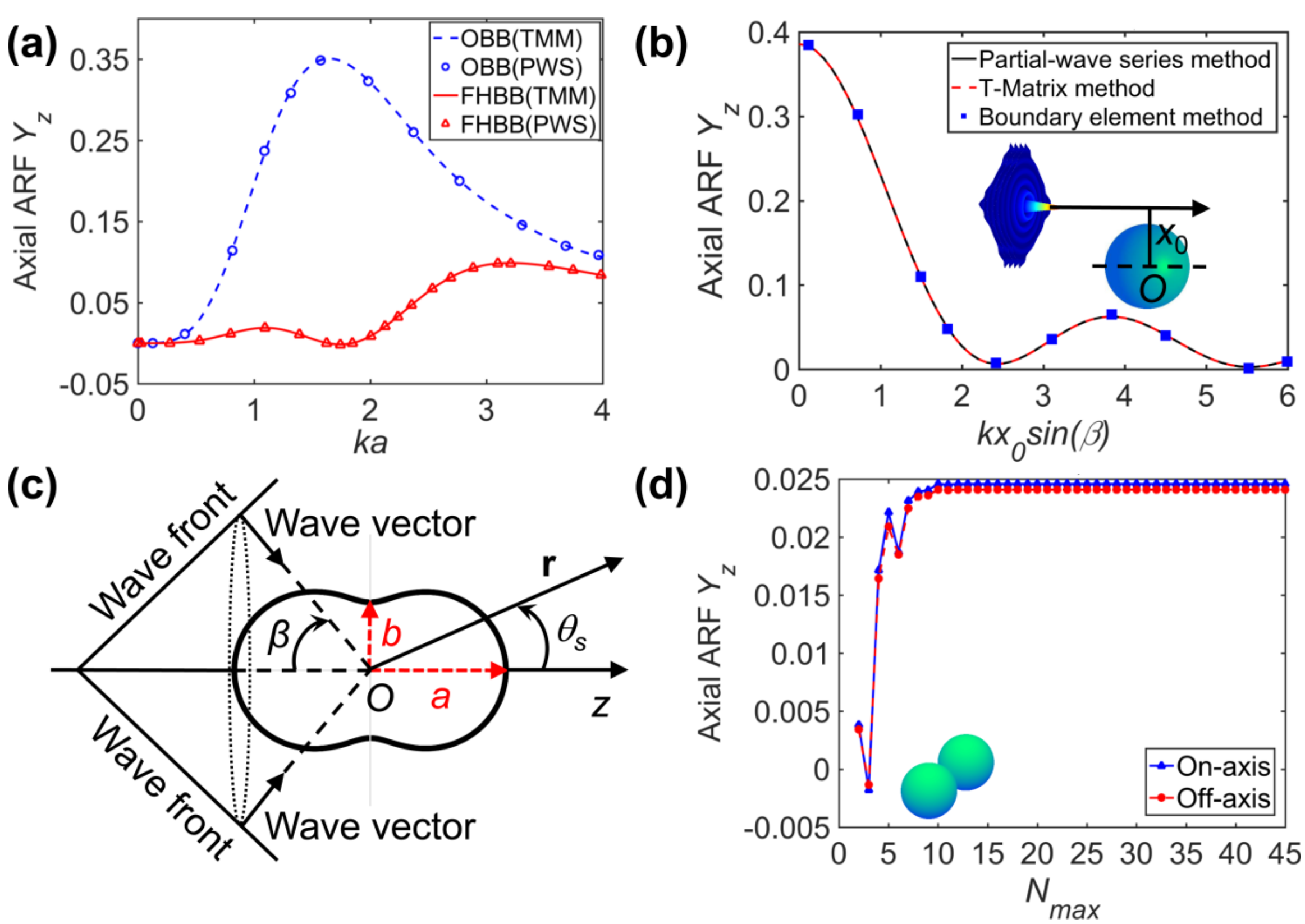

**Fig. 4**

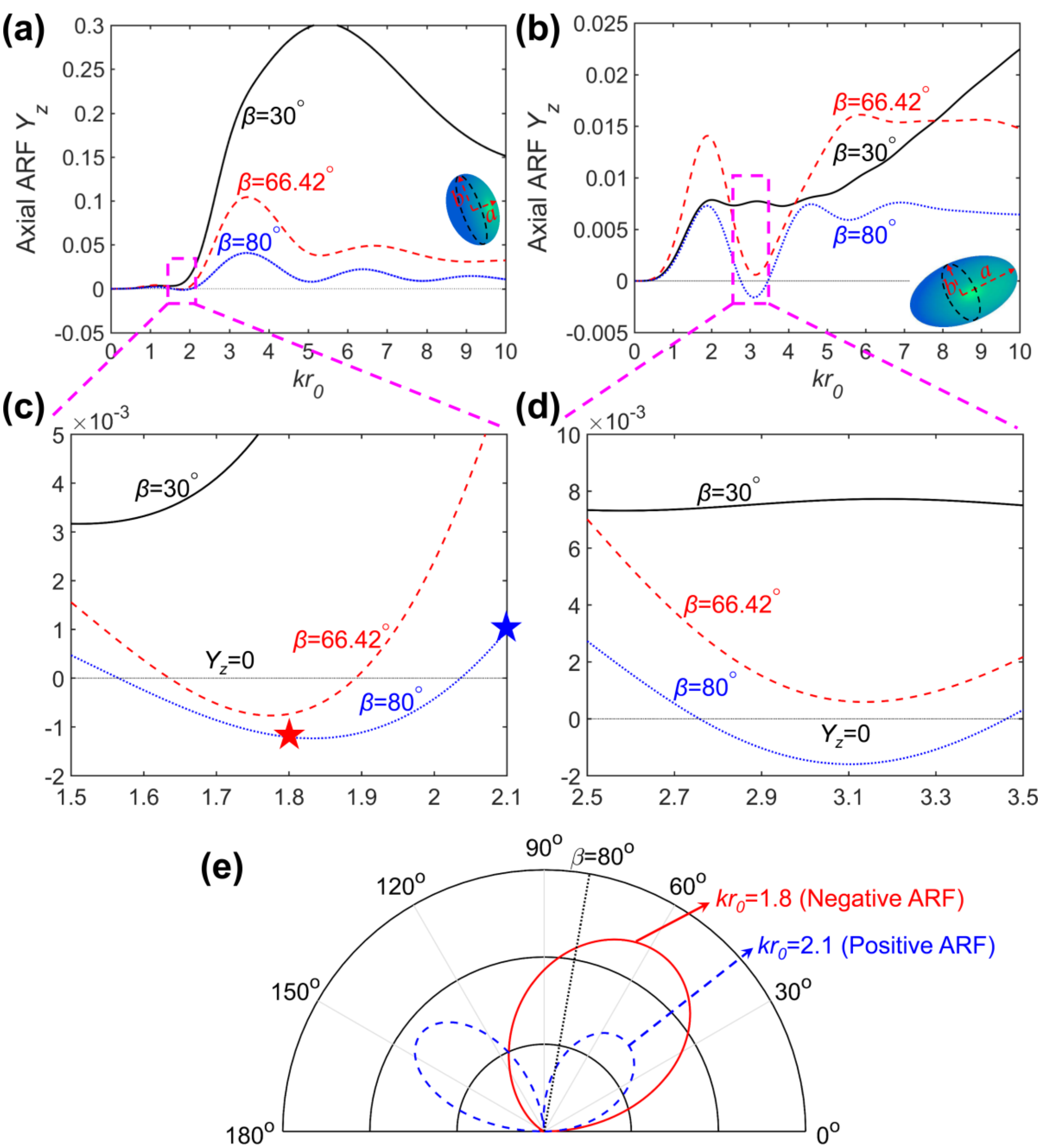

**Fig. 5**

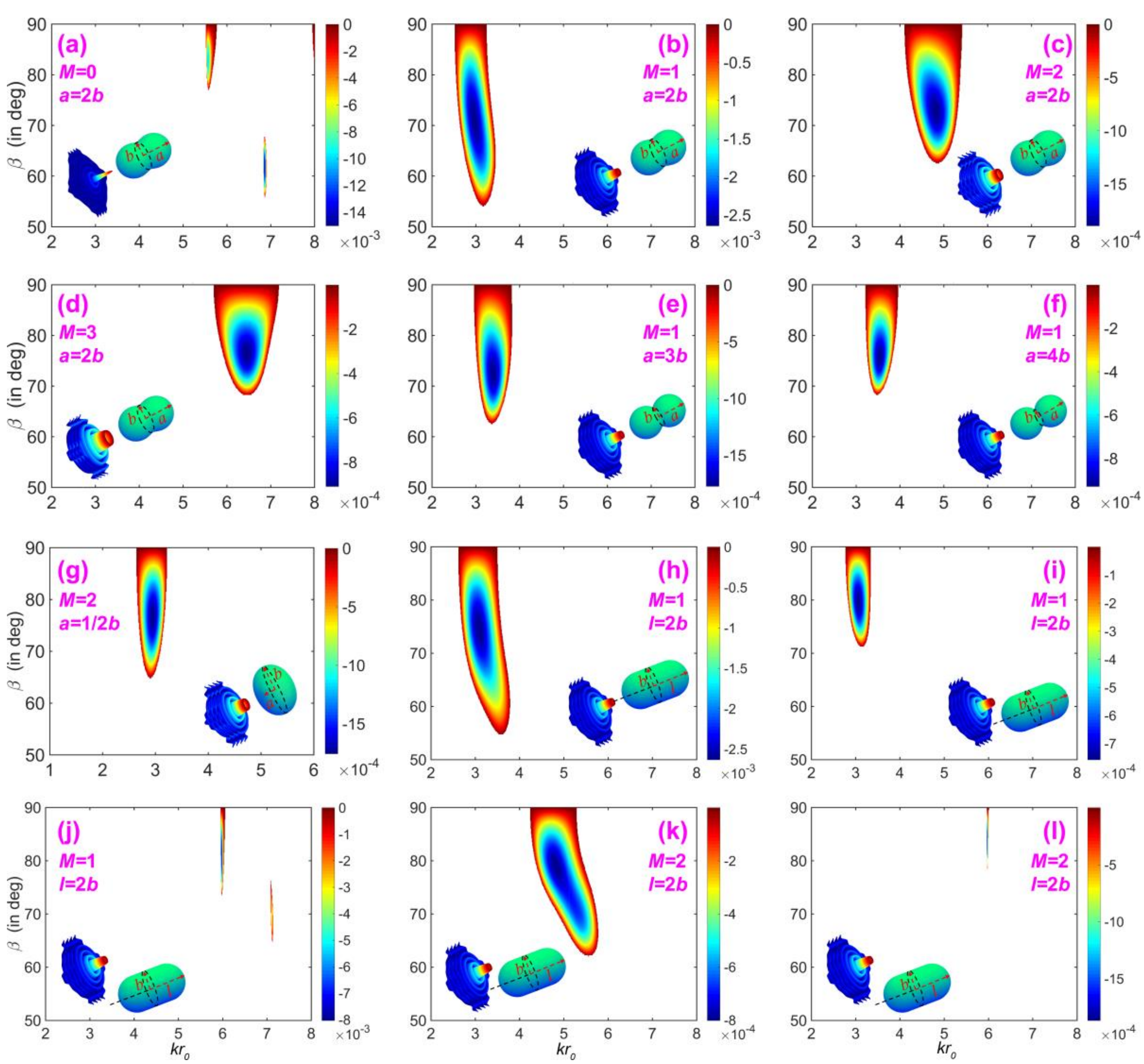

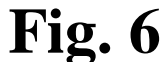

**Fig. 6**

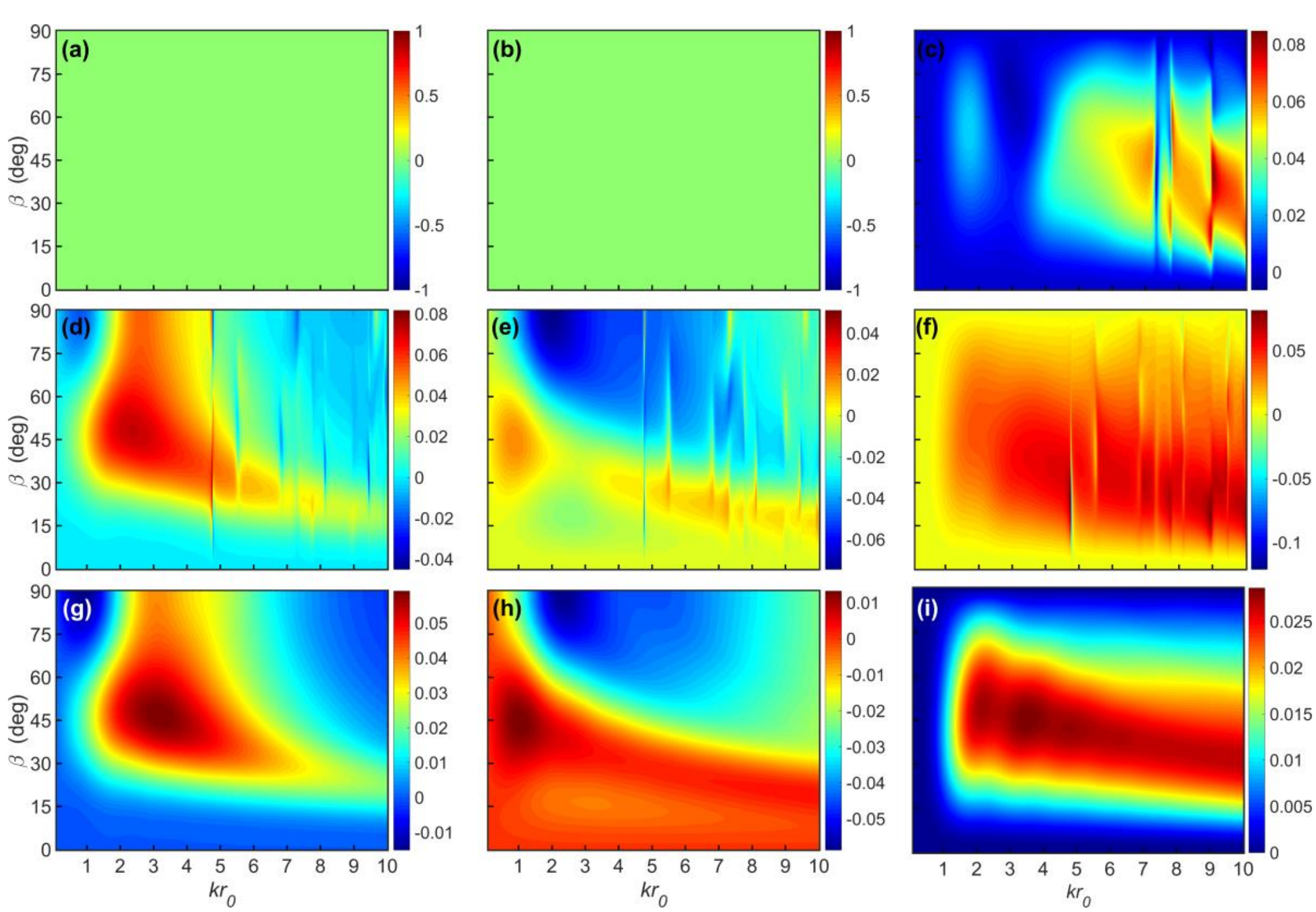

**Fig. 7**

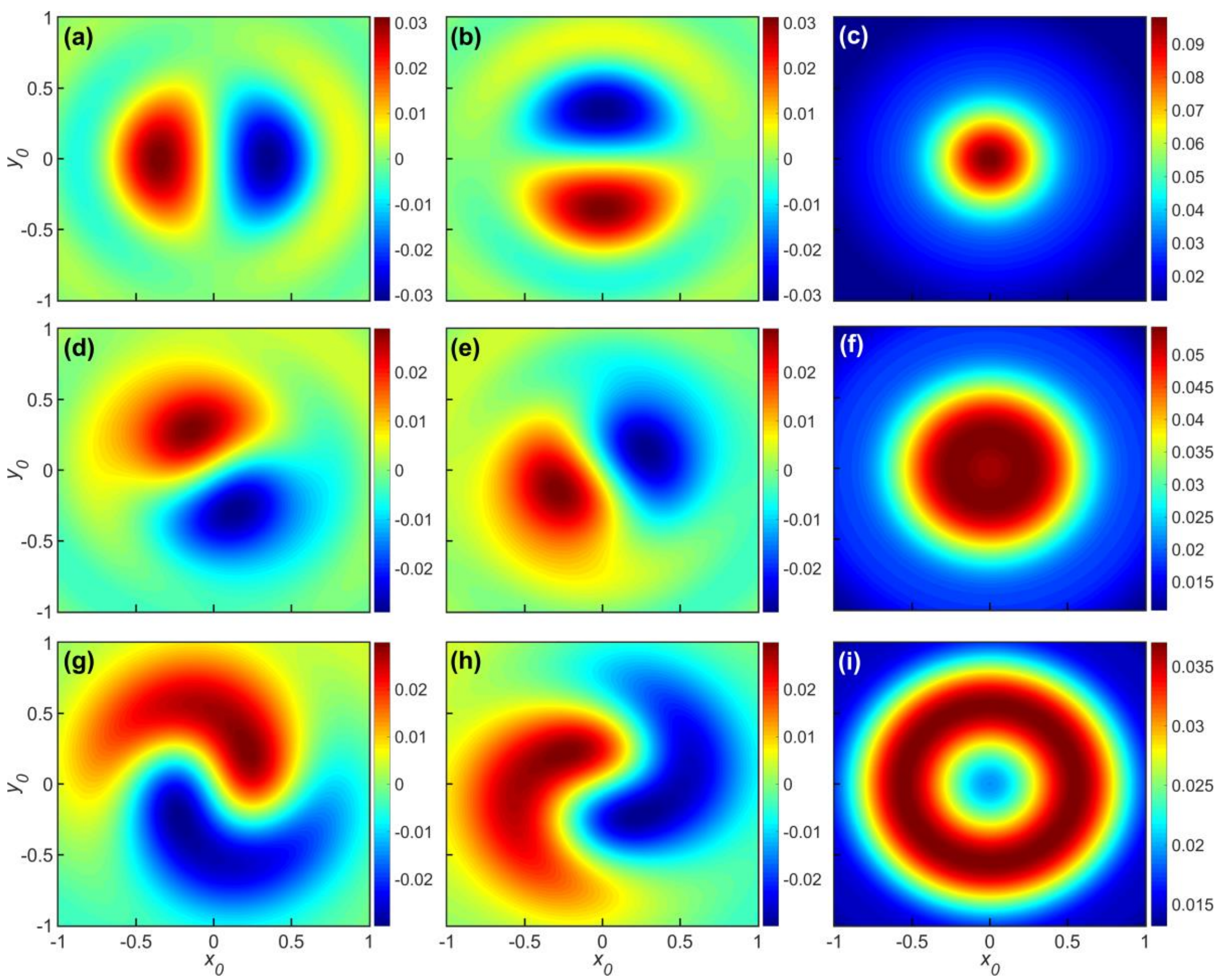

**Fig. 8**

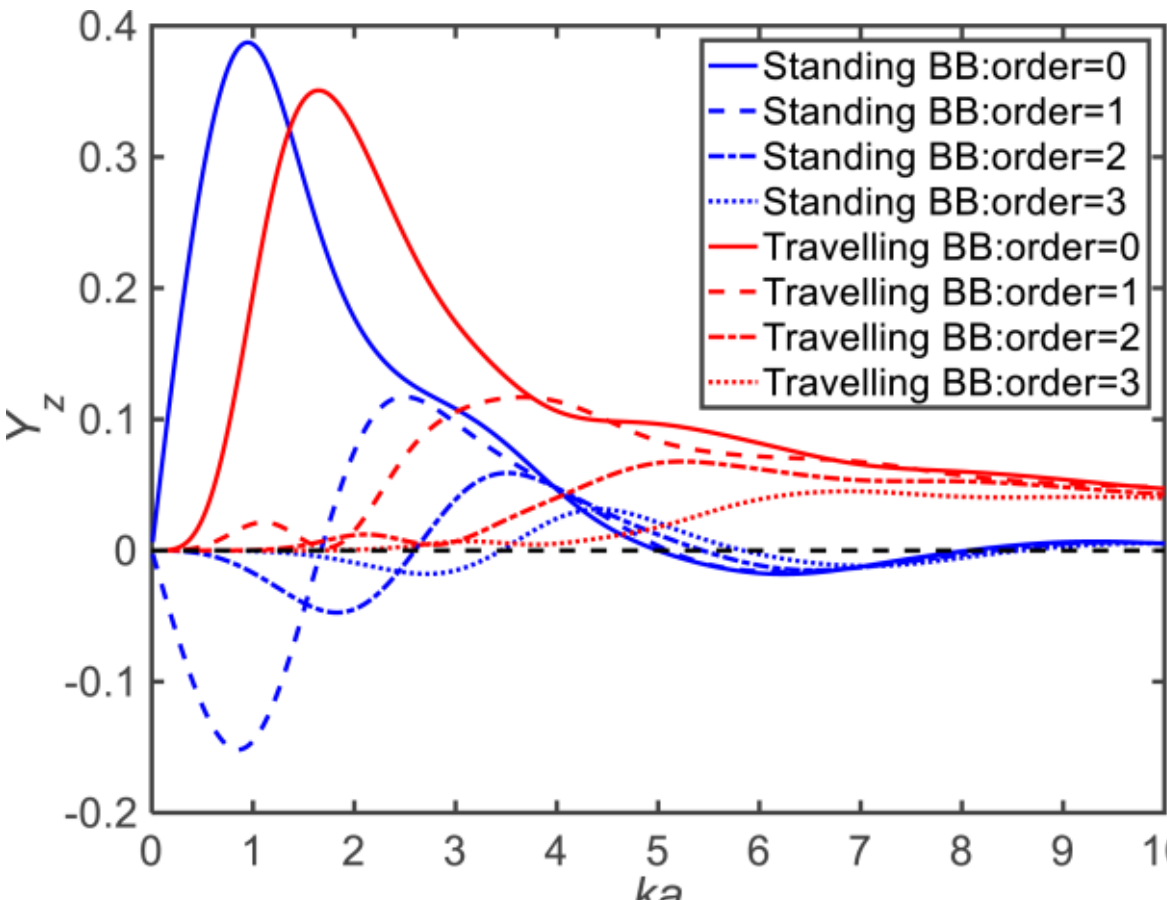